\documentclass[prl, twocolumn,superscriptaddress,longbibliography]{revtex4-1}

\usepackage{color}
\usepackage{graphicx}
\usepackage[colorlinks=true,citecolor=magenta,urlcolor=blue]{hyperref}
\usepackage{amsmath}
\usepackage{graphicx}
\usepackage{sistyle}
\usepackage{todonotes}
\usepackage{gensymb}
\usepackage{bbold}
\usepackage{bm}
\usepackage{natbib}
\usepackage{amsfonts}
\usepackage{dsfont}
\usepackage{empheq}
\usepackage{amssymb}
\usepackage{latexsym}
\usepackage{fancyhdr}
\usepackage[bbgreekl]{mathbbol}
\usepackage{epsfig}
\usepackage{color}

\usepackage{lipsum}
\usepackage{amssymb}

\usepackage{array}
\usepackage{tabulary}
\newcolumntype{K}[1]{>{\centering\arraybackslash}p{#1}}

\begin{document}

\title{24-Hour Relativistic Bit Commitment}
\author{Ephanielle Verbanis}
\author{Anthony Martin}
\email{anthony.martin@unige.ch}
\author{Rapha\"el Houlmann}
\author{Gianluca Boso}
\author{F\'elix Bussi\`eres}
\author{Hugo Zbinden}
\affiliation{Group of Applied Physics (GAP), University of Geneva, Chemin de Pinchat 22, CH-1211 Geneva 4, Switzerland}

\date{\today}

\begin{abstract}
Bit commitment is a fundamental cryptographic primitive in which a party
wishes to commit a secret bit to another party. Perfect security between
mistrustful parties is unfortunately impossible to achieve through the
asynchronous exchange of classical and quantum messages. Perfect security can nonetheless be achieved if each party splits into two agents exchanging classical information at times and locations satisfying strict relativistic constraints. A relativistic multi-round protocol to achieve this was previously proposed and used to implement a 2~millisecond commitment time. Much longer durations were initially thought to be insecure, but recent theoretical progress showed that this is not so. In this letter, we report on the implementation of a 24-hour bit commitment based on timed high-speed optical communication and fast data processing only, with all agents located within the city of Geneva. This duration is more than six orders of magnitude longer than before, and we argue that it could be extended to one  year and allow much more flexibility on the locations of the agents. Our implementation offers a practical and viable solution for use in applications such as digital signatures, secure voting and honesty-preserving auctions.

\end{abstract}
\pacs{}
\maketitle

Bit commitment is a cryptographic primitive in which a party Alice commits a secret bit to another party Bob, and later reveals it at a time of her choosing. A bit-commitment protocol is secure against a cheating Alice if it guarantees that she cannot reveal, without being caught, another bit than the one she initially committed to an honest Bob. The protocol is also secure against a cheating Bob if it guarantees that no information about the committed bit can be obtained before Alice reveals her commitment. Perfectly secure bit commitment between two mistrustful parties is known to be impossible through the asynchronous exchange of classical or quantum messages~\cite{Mayers1997,Lo1997,DAriano2007}. Arbitrarily long commitments using the exchange of quantum messages are possible if one makes the assumption that any quantum storage used by a dishonest party is bounded \cite{Damgard2005, Damgard2007} and noisy \cite{Wehner2008}. The security is however dependant on the parameter of the hardware (loss, detection noise, etc.) 

Alternatively, a scheme in which each party is split in two agents was shown to be secure against classical attacks under the assumption that no communication was possible between agents of the same party~\cite{Ben-Or1988}. Relativistic constraints on the communications between the different parties were suggested to enforce the no-communication assumption~\cite{Kent1999}, and a protocol using classical and quantum communication was proposed~\cite{Kent2012}. This protocol was experimentally realised and shown to be secure against classical and quantum attacks, even in the presence of unavoidable experimental imperfections~\cite{Lunghi2013,Yang2014}. These protocols are however limited to a single round of communication and therefore to a commitment of at most 21~ms if the agents are all located on Earth. 

To overcome this limitation, a new scheme using multiple rounds of classical communication was proposed and shown to be secure against classical attacks~\cite{Lunghi2015}. This scheme can in principle allow an arbitrary long commitment time by periodically sustaining the carefully timed classical communication between the two parties. The security analysis in Ref.~\cite{Lunghi2015} derives the following bound on the probability $\epsilon$ of a successful cheating attempt:
 \begin{equation} \label{old}
\epsilon \lesssim 2^{-n/2^{ (m-1)}},
\end{equation} where $n$ is the length of the bit string communicated between the agents of Bob and Alice at each round of the protocol, and $m+1$ is the number of rounds~\cite{Lunghi2015}. To keep $\epsilon \ll 1$, the length of the messages $n$ has to grow exponentially with the number of rounds. Therefore, an arbitrarily long commitment is in practice impossible to achieve. The implementation in Ref.~\cite{Lunghi2015} was limited to 6 rounds, yielding a 2~ms bit-commitment with agents separated by 131~km, i.e.~between Geneva and Bern. This could have been extended to a maximal value of 212~ms using antipodal locations on earth. 

Interestingly, the security bound of~\cite{Lunghi2015} was later improved significantly in two independent proofs considering classical attacks~\cite{Chakraborty2015,Fehr2015}. In both cases, the bound is linear in the number of rounds. For instance, the bound in \cite{Chakraborty2015} is
\begin{equation} \label{new}
\epsilon \leq m\, 2^{(-n+3)/2}.
\end{equation} 
These results open the way towards the implementation of much longer commitments. A more recent bound has been derived~\cite{pivoluska2016} showing a small reduction of the required resources.

In this Letter, we present the first experimental realisation of a 24-hour long bit commitment using agents suitably positioned 7~km apart, all within the immediate vicinity of Geneva. This increases the previous commitment time by more than six orders of magnitude compared to what was achieved previously using relativistic protocols~\cite{Lunghi2013,Lunghi2015}.

 We first describe the multi-round protocol and its security. The protocol contains a \textit{commit}, \textit{sustain} and \textit{reveal} phases. In the commit phase, the first agent of Bob $B_1$ sends a random $n$-bit string $x_1$ to the first agent of Alice $A_1$, who replies with the string $y_1=a_1$ to commit the bit~0, or $y_1=x_1\oplus a_1$ to commit the bit~1.
Here, "$\oplus$" is the bitwise XOR operation. The second agent $B_2$ starts the sustain phase by sending $x_2$ to agent $A_2$, who then returns $y_2=(x_2 \cdot a_{1}) \oplus a_2$, where "$\cdot$" is the multiplication in the Galois field $\mathbb{F}_{2^n}$. The third round is again between $B_1$ and $ A_1$, the fourth one between $B_2$ and $A_2$, and so on. At the k$^{\rm th}$ round, the agent $B_i$ sends $x_k$ and $A_i$ replies $y_k=(x_k \cdot a_{k-1}) \oplus a_k$, for $2\leq k \leq m$. Finally, to open the commitment, the following $A_i$ sends her commitment and $y_{m+1}=a_m$ to the corresponding $B_i$. With the set of questions $x_i$, the set of answers $y_i$ and the last string $a_m$, Bob can compute all the previous $a_k$ and verify that the bit he received during the reveal phase is the one committed by the first answer $y_1$ of agent $A_1$.  

\begin{figure}
\includegraphics[width=0.95\columnwidth]{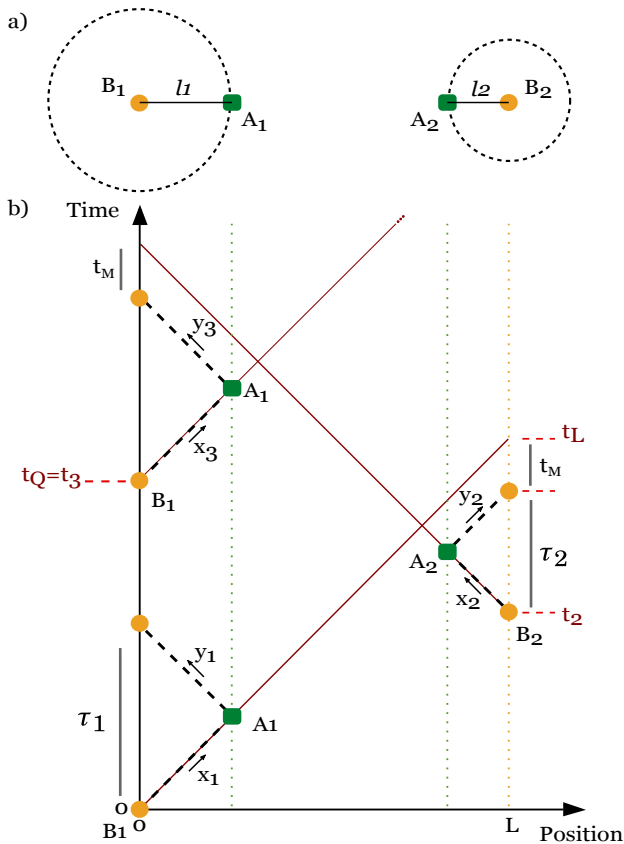}
\caption{(a) Positioning of the agents of Bob and Alice. Agent $B_i$ imposes that $A_i$ is within a distance $l_i$ from him and that she answers within a time $\tau_i$. (b) Space-time diagram of the multi-round protocol showing the relativistic constraints for the spatial configuration where $A_i$ is placed at the distance limit $l_i$. }
\label{fig_schema}
\end{figure}

 To comply with the relativistic constraints, Alice places one agent $A_i$ near each agent of Bob $B_i$. The distance $L$ between the agents of Bob is chosen such that by carefully timing the communication between $A_i$ and $B_i$, the security can be guaranteed. 
For our analysis, we consider a situation where Alice is allowed to place her agent $A_i$ at a maximum distance $l_i$ from their corresponding $B_i$, as shown in Fig. 1(a).  To ensure that no communication is possible between the two agents of Alice, we must impose the two following constraints: $A_i$ should answer before a time $\tau_i$ from the start of the corresponding round, or the protocol is aborted, and agent $B_i$ should start the round $t_L-(\tau_i+t_M)$ time after the start of the previous round, where $t_L$ is the time taken by light to travel the distance $L$ and $t_M$ is a chosen time margin. The security is guaranteed if $t_M$ is much greater than the timing uncertainties. Fig.~1(b) shows the relativistic constraints in the case where the agents of Alice are placed at the maximum allowed distance~$l_i$. An agent with a realistic data processing speed will have to place himself closer to answer within $\tau_i$. The time between the rounds of each agent of Bob, labelled $t_Q$, is related to $t_L$, $t_M$ and $\tau_i$ as 
$t_Q=2t_L-2t_M-(\tau_1+\tau_2)$. In terms of the distance parameters, this relation becomes: 
\begin{equation}
t_Q=\frac{2}{c}(L-(l_1+l_2))-2t_M.
\end{equation} The number of rounds is set by the spatial configuration, the chosen time margin $t_M$ and the time of commitment $T$. It can be related to $t_Q$ as:
\begin{equation}
m+1=\frac{2T}{t_Q}
\end{equation}

We now describe the implementation of the multi-round relativistic protocol. Each agent has a computer (Intel core i3 processor with 4 GB RAM) with a field-programmable gate array (FPGA) card (Xilinx SP605 evaluation board featuring Spartan-6 XC6SLX45T) installed to perform the computing tasks of the multi-round protocol. To ensure an accurate timing of exchange between $A_i$ and $B_i$, the FPGAs of Bob's agents are synchronized using local clocks. Specifically, each $B_i$ has access to an oven-controlled crystal oscillator (OCXO) with a nominal frequency of 10 MHz. This frequency is then multiplied inside the FPGA with a phase-locked loop to obtain a 125-MHz reference used to clock the computations steps. The OCXO are disciplined by a Global Positioning System (GPS) receiver, which provides a 1-PPS (one pulse per second) signal aligned to the \textit{Coordinated Universal Time} (UTC).
To make sure the FPGAs stay synchronized with the GPS reference, the 1-PPS signal from the receiver is also sent directly to the FPGA. An error of $\pm 1$ FPGA cycle (8 ns) on the expected $125\times 10^6$ cycles per second between two consecutive pulses from the GPS is tolerated. This uncertainty is sufficiently small to ensure that the relativistic constraints are satisfied.

 The agents $A_1$ and $B_1$ are located at the Group of Applied Physics (GAP) of the University of Geneva, while agents $A_2$ and $B_2$ are placed  7.0~km away from the GAP.  An optical signal in straight line would take 23.3 $\mu\text{s}$ to cover this distance. Before starting the protocol, the agents of Alice, and similarly the agents of Bob, share the appropriate size of random data to generate the strings $x_i$ and $a_i$. The strings to be communicated are transferred from the hard drive to the FPGA using a PCIe Gen1 x1 link. The maximum data transfer rate we can achieve with this system is 200 mega bytes per seconds (MBps). The exchange of the $n$-bit strings between $A_i$ and $B_i$ is performed via a 1~m optical link. We set the length of the strings to $n = 128$~bits.  With the various latencies in our system and the time needed to compute $y_k$, we can complete a round  in $1.8$ $\mu\text{s}$. To account for possible fluctuations in the duration and timing of the rounds, we impose that the agents of Alice answer within 3~$\mu\text{s}$ from the start of the round and take a margin $t_M$ of 3.3~$\mu\text{s}$, such that each round starts $6.3$ $\mu\text{s}$ before the earliest arrival time $t_L$ of the information from the previous round. Fixing $\tau_1=\tau_2=\tau= 3$~$\mu\text{s}$ corresponds to a situation where Bob imposes a maximum distance $l_1=l_2=l$ of 450~m to the agents of Alice. The shortest distance $L$ we could achieve with our system is limited by $\tau$ and $t_M$ to about 2.8~km.

We performed a 24-hour bit commitment with a security parameter $\epsilon$ of $7.8\times10^{-10}$, which requires $5\times10^9$ rounds, a total of 162~GB of data and an average transfer rate of 0.5~MBps. The amount of time needed to verify the commitment with the CPU of the computer is particularly long due to the computation of the Galois function. In our case, the computing was performed on a standard computer and took 72 hours, which is three times the commitment time $T$. This issue can be solved by using an FPGA to do the computing. We estimate that the verification time of our experiment could be lowered to about 90~minutes with a dedicated FPGA. Moreover, to shorten this time Bob could start the calculation for both bits in parallel of the bit commitment process and check at the end when Alice reveal the bit value.
\begin{table} [b]
{ \begin{tabular}{|c | c | c | c | c | c | c |} 
 \hline
  & $L$ [km] & $T$ & $\epsilon$ & $r$ [ Bps]  & Data [GB]& $T_{v}$ \\ \hline 
1 & 7.0 & 24 h & $7.8\times10^{-10}$ & $5\times10^5$  & 162 & 1 h 26 min  \\
& &  1 y & $2.8\times 10^{-7}$ & $5\times10^5$ & 59362& 530 h \\ \hline 
2 & 10000 & 24 h & $1\times10^{-12}$ & 649 &0.2 & 7 s \\
&  & 1 y & $3.9\times 10^{-10}$ & 649 & 81 & 44 min \\ \hline
\end{tabular}}
\caption{ Table showing the security parameter $\epsilon$, the data transfer rate $r$, the amount of data and the estimated verification time $T_v$ achievable with an FPGA for two different cases of spatial configurations and commitment times (24~h and 1~year). Case 1: $\tau=3$~$\mu\text{s}$, $l=450$~m and $t_M=3.3$~$\mu\text{s}$; Case 2: $\tau=20$~ms, $l=3000$~km and $t_M=1$~ms . The length of the communicated string is set to 128 bits in both cases.}
\label{table}
\end{table}

The protocol could in practice be achieved in a large range of spatial configurations. Table~1 shows the required resources and the security parameter for two different spatial configurations and commitment times. The case 1 corresponds to our experiment, i.e.~Bob's agents located seven kilometres apart and Alice's agents placed near the agents of Bob. Bit commitments in this regime are the most demanding in terms of resources. Increasing the distance between the agents of Bob effectively reduces the number of rounds, hence reduces the resources needed. The case 2 corresponds to a situation where the agents of Bob are separated by a straight line distance of 10,000 km and impose that the agents of Alice answer within 20 ms from the beginning of the rounds. In 20~ms, light covers a straight line distance of 6000~km. This means that Alice must position her agents within a 3000~km radius from Bob's agents. In practice, this radius will be smaller, depending on the speed and latency of the communication link between the agents.  As an example, Bob could have one agent in the center of the United States of America and one agent in the center of China, and could engage the commitment with agents of Alice located within a large area of these two countries. Indeed, we can imagine networks formed by several interconnected nodes allowing commitments from parties placed all over the world.

The security against a cheating Alice relies on the correct timing of each round with respect to the previous one. In order to exclude attacks on the timing, such as spoofing of the GPS signal, Bob could rely solely on local clocks, provided they have been synchronized before, e.g. with a third clock travelling between the agents. For a margin of 1~ms (case~2), the uncertainty on the frequency of the clocks must be less than~$1.2\times10^{-8}$ for a commitment time of 24~hours and less than $3.1\times10^{-11}$ for a commitment time of 1~year, which can be achieved with commercially available clocks\cite{Diddams2004}.

Our implementation demonstrates long commitment times without the need of large computing power in realistic and potentially useful scenarios. The implementation offers a practical and viable solution for use in applications such as coin tossing~\cite{Blum1981}, zero-knowledge proofs~\cite{Damgard1999}, digital signature~\cite{Boneh2000} and secure multi-party computations, including secure voting\cite{Broadbent2008} and honesty-preserving auctions~\cite{Boneh2000}.

\begin{acknowledgments}
We thank Jedrzej Kaniewski for helpful discussions. This work was supported financially by the Swiss NCCR-QSIT.
\end{acknowledgments}

\bibliography{bitcom}

\end{document}